\documentclass {aa}
\usepackage {amssymb}
\usepackage {times}
\usepackage {mathptm}
\usepackage {graphics}
\usepackage {subfigure}
\usepackage[dvips] {epsfig}
\usepackage {natbib}
\bibpunct {(} {)} {;} {a} {} {,}

\newcommand {\rb}[1] {\raisebox {1.5ex}[-1.5ex] {#1}}

\def\Teff  {T$_ {\mbox {\scriptsize eff}}$}
\def\logg  {$\log g$}
\def\vt  {$v_ {t}$}
\def\kms  {$ {\rm km \: s^ {-1}}$}
\def\dble  {HE\,1353--2735}
\def\sgle  {HE\,1303--2708}
\def\Ha  {H$\alpha$}

\begin {document}     
\thesaurus {08(08.01.1; 08.02.4; 08.06.3; 08.09.2; 08.19.2; 04.19.1)}

\title {Abundance analysis of two extremely metal-poor stars
  from the Hamburg/ESO Survey\thanks {Based on observations collected at the
    European Southern Observatory, Paranal, Chile}}

\author {
E. Depagne\inst {1} \and
V. Hill\inst {2} \and 
N. Christlieb\inst {3} \and
F. Primas\inst {2}
}

\institute {
  Observatoire de Paris-Meudon, F-92125 Meudon Cedex, 
  France, Eric.Depagne@obspm.fr
  \and European Southern Observatory, D-85748 Garching, Germany,
  [vhill,fprimas]@eso.org
  \and Hamburger Sternwarte, Gojenbergsweg 112,
  D-21029 Hamburg, Germany, nchristlieb@hs.uni-hamburg.de
}

\offprints {Eric.Depagne@obspm.fr}
\date {Received 9-Aug-2000; accepted ??}
\titlerunning {Two metal-poor stars from the Hamburg/ESO Survey}
\authorrunning {Depagne et al.}
\maketitle

\begin {abstract}
  We report on the first high spectral resolution analysis of extremely
  metal-poor halo stars from the Hamburg/ESO objective-prism survey (HES).
  The spectra were obtained with UVES at VLT-UT2. The two stars under
  investigation (\object {HE\,1303--2708} and \object {HE\,1353--2735}) are
  main-sequence turnoff-stars having metal abundances of $\mbox {[Fe/H]}=-2.85$
  and $-3.20$, respectively. The stellar parameters derived from the UVES
  spectra are in very good agreement with those derived from
  moderate-resolution follow-up spectra.  HE\,1353--2735 is a double-lined
  spectroscopic binary.
  
  The two stars nicely reproduce the strong scatter in [Sr/Fe] observed for
  extremely metal-poor stars. While we see a strong Sr~II $\lambda
  4215$\, {\AA} line in the spectrum of HE\,1303--2708
  ($\mbox {[Sr/Fe]}=-0.08$), we can only give an upper limit for HE\,1353--2735
  ($\mbox {[Sr/Fe]}<-1.2$), since the line is not detected.
  We report abundances of Mg, Ca, Sc, Ti, Cr for both stars, and Co, Y for         HE\,1303--2708 only. These abundances do not show any abnormalities with
  respect to known trends for metal-poor stars.Lithium is also detected in these   stars, to a level which places them among Lithium-plateau metal-poor dwarfs. 
\end {abstract}


\section {Introduction}
The most metal-poor stars carry the fossil record of the chemical compostion of the Galaxy, and hence allow one to study the ealiest epochs of Galactic chemical evolution. Moreover, there are also \emph {cosmological} applications for
metal-poor stars, e.g. the determination of the primordial Lithium abundance
\citep {Spite/Spite:1982,Ryanetal:2000}, constraining the baryon density
parameter, or individual age determinations \cite [e.g.,][] {Cowanetal:1999},
thus setting a lower limit to the age of the Universe. Therefore, in the past
decade there has been a fast-growing interest in these stars within the
astronomical community. With VLT-UT2 and UVES now being in operation, it has
become feasible to study large samples of metal-poor stars at high resolution
and high $S/N$ in a very reasonable time.

For the past decade, the major source of metal-poor stars has been the so-called HK survey of Beers and collaborators \citep {BPSI,BPSII,TimTSS} . Now, an even larger survey volume within which to search for metal-deficient stars has been opened up by use of the digital objective-prism spectra of the Hamburg/ESO Survey \citep [HES;][] {HESpaperI,heshighlights,HESpaperIII}, since the HES is more than one magnitude deeper than the HK survey. \cite {Christlieb:2000} has shown that the selection of metal-poor candidates in the HES by automatic spectral classification is much more efficient than the manual selection applied to the HK survey, so that the amount of telescope time needed for follow-up spectroscopy is reduced, and a \emph {full} exploitation of the $\sim 8\times$ larger HES volume (compared to the HK survey alone) becomes feasible. An alternative selection method in the HES is the so-called Ca~K index method \citep {nctss}, which looks for stars characterized by a Ca\,II~K line weaker than expected for their $B-V$ color. This method is almost as efficient in selecting metal-poor stars as automatic classification \citep {Christliebetal:2000}.

In this paper we report on the first high-resolution observations of
metal-poor stars from the HES. Two stars, \object {HE\,1303--2708} and
\object {HE\,1353--2735}, were observed with VLT-UT2 in the course of UVES
Science Verification in February 2000.

\section {HES candidate selection and spectroscopic follow-up}

\begin {figure*}[htbp]
  \begin {center}
    \leavevmode
    \epsfig {file=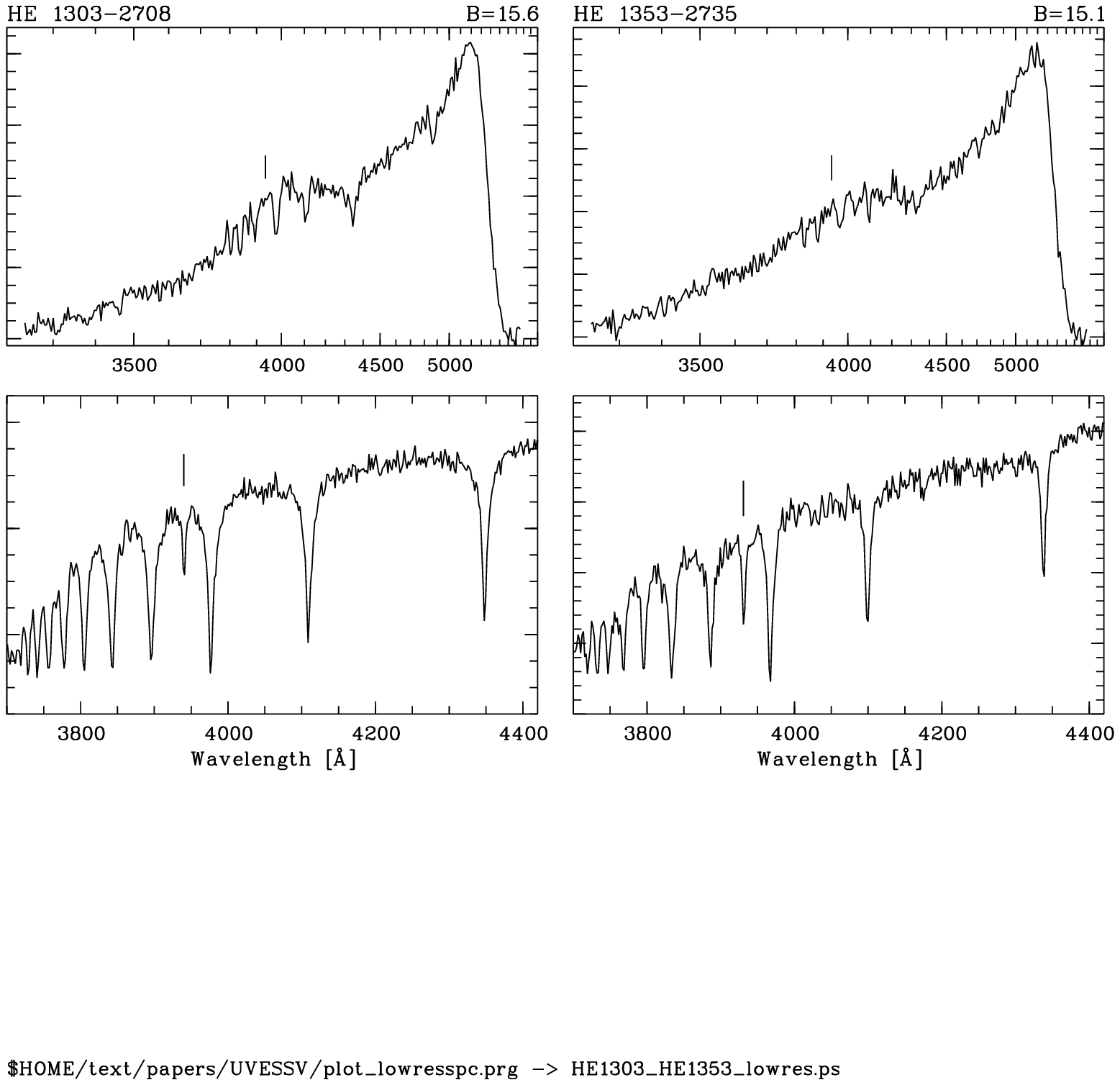, clip=, width=17.5cm,
      bbllx=81, bblly=420, bburx=531, bbury=734}
  \end {center}
  \caption {\label {Fig:lowres_spectra} Upper panels: HES objective-prism
    spectra of \object {HE\,1303--2708} and \object {HE\,1353--2735}; lower
    panels: Moderate-resolution spectra obtained with the ESO-NTT and EMMI.
    The position of Ca~K is marked. The ordinates are photographic density in
    arbitrary units (upper panels), and relative flux (lower panels).}
\end {figure*}

The two stars described in this paper were selected by the Ca~K-index method
\citep {nctss}. As is obvious from Fig.  \ref {Fig:lowres_spectra},
\object {HE\,1303--2708} and \object {HE\,1353--2735} were both selected because
they show \emph {no} Ca\,II~K ($\lambda\,3933$\, {\AA}) line in the HES spectra.

Tab. \ref {Tab:HESCoordPhot} lists coordinates (derived from the DSS-I, see
e.g. \texttt {http://archive.eso.org/dss/dss}), and photometry for the two
stars.

\begin {table*}[htbp]
  \begin {center}
    \caption {\label {Tab:HESCoordPhot}  Coordinates, photometry and
    stellar parameters for the two metal-poor stars. Photometry was
    derived directly from the HES plates \citep[for the calibrations
    see][] {Christlieb:2000}, with the exception of $B-V$ for \dble,
    which was measured in April 2000 with NTT/EMMI. Stellar parameters were
    derived from moderate-resolution follow-up spectra. In the second row
    for each star we list the $1\sigma$ errors.}
    \begin {tabular} {llllllllllll}
      \hline\hline Name & $\alpha (2000)$ & $\delta (2000)$ &
      $B_J$ & $V$ & $B-V$ & $U-B$ & $c_1$ & $T_ {\mbox {\scriptsize eff}}$ & $\log g$ &
      $\mbox {[Fe/H]}_ {\mbox {\scriptsize HES}}$ &
      $\mbox {[Fe/H]}_ {\mbox {\scriptsize HK}}$
      \rule {0.0ex} {2.3ex}\\[0.2ex]\hline\\[-2.2ex]
      HE\,1303--2708 & 13 06 37.8 & $-$27 24 22 & 15.5 & 15.3 & 0.45 & $-0.18$ & 0.49
 & 6500\,K & 4.2 & $-2.8$ & $-3.3$\\
      & & & \hspace {1ex}0.2 & \hspace {1ex}0.2 & 0.03 &  \hspace {1.85ex}0.09 & 0.15
 & \hspace {1ex}200\,K & 0.5 & \hspace {1.85ex}0.2 & \hspace {1.85ex}0.2\\[0.8ex]
      HE\,1353--2735 & 13 56 42.5 & $-$27 49 54 & 15.0 & 14.7 & 0.38 & $-0.27$ & 0.23
 & 6000\,K & 4.8 & $-2.9$ & $-3.4$\\
      & & & \hspace {1ex}0.2 & \hspace {1ex}0.2 & 0.07 &  \hspace {1.85ex}0.09 & 0.15
 & \hspace {1ex}200\,K & 0.5 & \hspace {1.85ex}0.2 & \hspace {1.85ex}0.2\\
\hline\hline
    \end {tabular}
  \end {center}
\end {table*}


Spectroscopic follow-up observations were obtained in April 1999 using EMMI
attached to the ESO-NTT. From these moderate-resolution ($\sim 5\,$ {\AA}
FWHM), but good quality spectra (see Fig. \ref {Fig:lowres_spectra}), having
$S/N\sim 50$ at Ca\,II~K and $S/N>60$ at the Mg~I\,b triplet
($\lambda\lambda\, 5167,5173,5184$\, {\AA}), the stellar parameters listed in
Tab. \ref {Tab:HESCoordPhot} were derived via the spectrum synthesis technique.
Ca~K, Mg~I\,b and Balmer line wings were used as indicators for metallicity,
gravity and effective temperature, respectively. Plane-parallel LTE model
atmospheres of Reetz et al. (1999, priv. comm.) were used. The analysis was
performed strictly differentially to the Sun, and the well-known
Calcium-overabundance of $\sim 0.4$\,dex for metal-poor stars \citep[see
e.g.][] {Wheeleretal:1989} was taken into account.


By re-observing and re-analysing a couple of metal-poor stars from the HK
survey (Beers 1999, priv. comm.), it turned out that there is an average
difference of $0.5$\,dex between the HES and HK survey [Fe/H] scales
\citep {Christlieb/Beers:2000}, in the sense that stars appear to be more metal-poor
in the HK survey than in the HES. Unfortunately it is not possible to identify
the physical reason for this difference, because \cite {HKrecalib} use average
[Fe/H] values from the literature (which were hence determined by using many
different model atmospheres) to calibrate their method. In order to make the
HES metallicities comparable with those of \cite {HKrecalib}, we also list
scaled [Fe/H] values in Tab. \ref {Tab:HESCoordPhot}.

\section  {UVES observations}
 
The observations were performed from February 10 to 18, 2000, as part of the
VLT-UT2 Science Verification of UVES, the UV and Visual Echelle Spectrograph
recently commissioned \citep {dodorico00}. The data from this program were made
public to the ESO community in April 2000 (see
\texttt {www.eso.org/science/ut2sv/UVES\_SV.html}). The spectrograph setting
(Dichroic mode, central wavelength 4700\, {\AA} in the blue arm, and
7850\, {\AA} in the red arm) provided a wavelength coverage of
4100--5300\, {\AA} and 6000--9700\, {\AA}, and the entrance slit of 1$\arcsec$
yielded a resolving power of R$\sim$45\,000. Table \ref {T-log} gives a short
log of the observations, together with the achieved $S/N$.

\begin {table}
\caption {\label {T-log} Log-book of the UVES observations. $S/N$ refers
  to the signal-to-noise ratio per 0.025\, {\AA} pixel, at 5200\, {\AA}
  and 6700\, {\AA} for the blue and red spectra, respectively.}
\begin {flushleft}
  \begin {tabular}  {lcccr}
    \noalign {\smallskip}\hline\hline
    \noalign  {\smallskip}
    Star  & Setting & $V$\,[mag] & $t$\,[min] & \multicolumn {1} {c} {$S/N$} \\\hline
    \noalign  {\smallskip}
    HE\,1303--2708 & B470 &           & 225 & 70\\
    HE\,1303--2708 & R785 & \rb {15.3} & 290 & 150\\[0.8ex]
    HE\,1353--2735 & B470 &           & 180 & 55\\
    HE\,1353--2735 & R785 & \rb {14.7} & 180 & 120\\
    \noalign  {\smallskip}\hline\hline
  \end {tabular}
\end {flushleft}
\end {table}

\begin {figure}[htbp]
\begin {flushleft}
        \centering      
        \resizebox  {\hsize} {!}
         {\includegraphics  {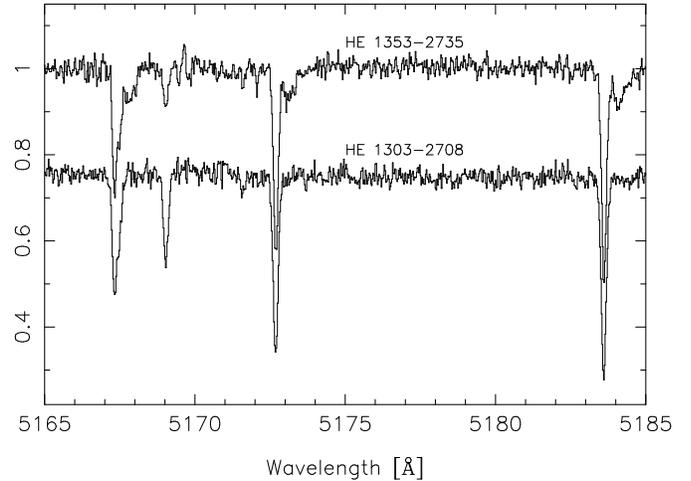} }
\caption {Mg\,I triplet region for both HES stars. Note that
 {\dble} is a double-lined binary, as can be seen from the faint lines
displaced towards the red of the main lines of the spectrum. The spectrum
of  {\sgle} was shifted vertically by 0.2 units in relative flux.}\label {F-mg}
\end {flushleft}
\end {figure}

The spectra were reduced using the UVES context within MIDAS, which performs
bias subtraction (object and flat-field), inter-order background subtraction
(object and flat-field), optimal extraction of the object (above the sky,
rejecting cosmic ray hits), division by a flat-field frame extracted with the
same weighted-profile as the object, wavelength calibration and rebinning to a
constant step and merging of all overlapping orders. The spectra were then
corrected for barycentric radial velocity and co-added, and finally normalised.
Fig. \ref {F-mg} is an example of the reduced spectrum around the magnesium
triplet for both stars.

As can be noted in that figure, it turned out that  {\dble} is a double-lined
bi\-nary with a faint system of lines shifted to the red by $\sim 25$\,\kms.
To serve as first epoch for future follow-up of the system, we note that the
radial velocities observed for the two components are $v_ {\mbox {\scriptsize
    rad,A}}=-213.8$\,\kms and $v_ {\mbox {\scriptsize rad,B}}=-187.7$\,\kms, at
$\mbox {MJD}=51584.31030$.

\section  {Stellar parameters and abundances}

\begin  {table}
\caption {\label {T-lines} Line by line abundances for all elements observed in
         {\sgle} and \dble, except iron. We do not list equivalent widths for
        \dble, since the abundances for this star were derived from spectrum
        synthesis.} 
        \begin  {flushleft}
                \begin  {tabular}  {lllllll}
\hline\hline
Element & $\lambda$ & $\chi_ {ex}$ & $\log$ gf & W$_ {1303}$&
        $\varepsilon_ {1303}$ & $\varepsilon_ {1353}$ \\
 & [\AA] & [eV] &  & [m\AA]\\ 
\hline
Li\,I & 6707.766 & 0.00 & \hspace {1.85ex}0.00 &      & 2.12 & 2.06 \\
 & 6707.916 & 0.00 & $-$0.30 &     & 2.12 & 2.06 \\
\hline
Mg\,I & 4702.998 & 4.34 & $-$0.37 & 13.6 &  4.86 & \\
 & 5167.360 & 2.71 & $-$0.86 & 74.4 &  5.19 & 4.8 \\
 & 5172.688 & 2.71 & $-$0.38 & 85.8 &  4.92 & 4.8 \\
 & 5183.603 & 2.72 & $-$0.16 & 95.8 &  4.90 & 4.8 \\
\hline
Ca\,I & 4226.740 & 0.00 & $-$3.36 &      &       & 3.5 \\
 & 4289.372 & 1.88 & $-$0.30 &  5.9 &  3.71 & \\
 & 4302.526 & 1.90 & \hspace {1.85ex}0.28 & 20.3 &  3.75 & \\
 & 4318.649 & 1.90 & $-$0.21 &  7.4 &  3.74 & \\
 & 4434.951 & 1.89 & $-$0.01 & 12.8 &  3.78 & \\
 & 4454.769 & 1.90 & \hspace {1.85ex}0.26 & 20.4 &  3.77 & 3.1 \\
\hline
Sc\,II & 4246.823 & 0.31 &  \hspace {1.85ex}0.32 & 27.3 &  0.40 & \\
 & 4314.080 & 0.62 & $-$0.10 & 12.1 &  0.64 & \hspace {-1.85ex}$-$0.2\\
 & 4320.720 & 0.61 & $-$0.26 &  6.4 &  0.49 & \\
\hline
Ti\,II & 4290.213 & 1.16 & $-$1.10 & 15.3 &  2.63 & \\
 & 4300.042 & 1.18 & $-$0.75 & 29.9 &  2.70 & 2.2 \\
 & 4301.908 & 1.16 & $-$1.43 &  9.2 &  2.72 & \\
 & 4395.033 & 1.08 & $-$0.65 & 37.9 &  2.68 & 2.2 \\
 & 4399.754 & 1.24 & $-$1.27 &  9.3 &  2.62 & \\
 & 4417.716 & 1.16 & $-$1.42 & 12.5 &  2.84 & \\
 & 4443.800 & 1.08 & $-$0.81 & 25.8 &  2.55 & 2.1 \\
 & 4468.492 & 1.13 & $-$0.77 & 29.1 &  2.64 & 1.8 \\
 & 4501.274 & 1.12 & $-$0.86 & 28.4 &  2.70 & 2.0 \\
 & 4533.957 & 1.24 & $-$0.76 & 26.6 &  2.66 & 2.2 \\
 & 4549.614 & 1.58 & $-$0.47 &      &   & 2.3 \\
 & 4563.753 & 1.22 & $-$0.95 & 23.0 &  2.74 & 2.2 \\
 & 4571.969 & 1.57 & $-$0.52 & 27.4 &  2.74 & 2.1 \\
\hline
Cr\,I & 4254.333 & 0.00 & $-$0.11 & 24.3 &  2.58 & 2.2 \\
 & 4274.793 & 0.00 & $-$0.23 & 19.1 &  2.55 & 2.2 \\
 & 4289.712 & 0.00 & $-$0.36 & 13.1 &  2.48 & 2.2 \\
 & 5206.061 & 0.94 & \hspace {1.85ex}0.02 & 15.7 &  2.97 & \\
 & 5208.400 & 0.94 & \hspace {1.85ex}0.16 & 16.1 &  2.84 & 2.2 \\
\hline
Co\,I & 4121.308 & 0.92 & $-$0.32 &  8.9 &  2.72 & \\
\hline
Sr\,II & 4215.520 & 0.00 & $-$0.17 & 40.5 & \hspace {-1.85ex}$-$0.03 &\hspace {-2.25ex}$<-1.5 $ \\
\hline
 Y\,II & 5087.367 & 1.08 & $-$0.17 &  2.6 &  0.21 & \\
\hline
Ba\,II & 4554.045 & 0.00 & \hspace {1.85ex}0.17 & 15.2 & \hspace {-1.85ex}$-$0.75 &\hspace {-2.5ex}$<-1.5$\\
 & 4934.080 & 0.00 & $-$0.16 &  9.4 & \hspace {-1.85ex}$-$0.70 & \\
\hline\hline
                \end  {tabular}
        \end  {flushleft}
\end  {table}
                
\begin  {table}
\caption  {Iron abundances in  {\sgle} and \dble.}\label {T-linesFe}
        \begin  {flushleft}
                \begin  {tabular}  {lllllll}
\hline\hline
Element & $\lambda$ & $\chi_ {ex}$ & $\log$ gf & W$_ {2708}$&
$\varepsilon_ {2708}$ & $\varepsilon_ {2735}$ \\       
 & [\AA] & [eV] &  & [m\AA]\\ 
\hline
Fe\,I & 4118.545 & 3.57 & \hspace {1.85ex}0.21 & 15.9 &  4.90 & \\
 & 4132.052 & 1.61 & $-$0.67 & 33.5 &  4.55 & 4.5 \\
 & 4143.403 & 3.05 & $-$0.20 &  8.0 &  4.52 & \\
 & 4143.866 & 1.56 & $-$0.51 & 42.9 &  4.57 & 4.5 \\
 & 4181.736 & 2.83 & $-$0.37 & 11.0 &  4.65 & 4.3 \\
 & 4187.040 & 2.45 & $-$0.55 & 13.6 &  4.61 & 4.5 \\
 & 4187.792 & 2.43 & $-$0.55 & 16.1 &  4.68 & 4.3 \\
 & 4191.425 & 2.47 & $-$0.67 &  8.2 &  4.50 & 4.3 \\
 & 4199.090 & 3.05 & \hspace {1.85ex}0.16 & 17.4 &  4.55 & 4.3 \\
 & 4202.024 & 1.48 & $-$0.71 & 37.9 &  4.58 & 4.5 \\
 & 4210.321 & 2.48 & $-$0.93 &  4.0 &  4.43 & \\
 & 4227.430 & 3.33 & \hspace {1.85ex}0.27 & 12.4 &  4.50 & \\
 & 4233.582 & 2.48 & $-$0.60 & 15.2 &  4.74 & 4.2 \\
 & 4235.935 & 2.43 & $-$0.34 & 17.6 &  4.52 & 4.3 \\
 & 4250.112 & 2.47 & $-$0.41 & 18.1 &  4.63 & 4.3 \\
 & 4250.782 & 1.56 & $-$0.71 & 33.9 &  4.55 & 4.5 \\
 & 4260.476 & 2.40 & \hspace {1.85ex}0.08 & 35.8 &  4.54 & 4.4 \\
 & 4271.149 & 2.45 & $-$0.35 & 18.6 &  4.57 & 4.3 \\
 & 4271.757 & 1.49 & $-$0.16 & 65.8 &  4.72 & 4.5 \\
 & 4282.405 & 2.18 & $-$0.78 & 11.4 &  4.51 & 4.4 \\
 & 4294.110 & 1.49 & $-$1.11 & 35.9 &  4.92 & 4.3 \\
 & 4299.230 & 2.43 & $-$0.41 & 16.4 &  4.54 & 4.4 \\
 & 4307.896 & 1.56 & $-$0.07 & 74.3 &  4.91 & 4.3 \\
 & 4315.104 & 2.20 & $-$0.97 &      &   & 4.3 \\
 & 4325.760 & 1.61 & \hspace {1.85ex}0.01 & 68.0 &  4.70 & 4.5 \\
 & 4375.921 & 0.00 & $-$3.03 &  8.9 &  4.70 & 4.3 \\
 & 4383.545 & 1.48 & \hspace {1.85ex}0.20 &  7.5 &  4.65 & 4.5 \\
 & 4404.746 & 1.56 & $-$0.14 &  0.2 &  4.59 & 4.3 \\
 & 4415.118 & 1.61 & $-$0.61 &  7.6 &  4.57 & 4.5 \\
 & 4427.305 & 0.05 & $-$2.92 &  6.7 &  4.50 & 4.4 \\
 & 4461.675 & 0.09 & $-$3.21 &      &   & 4.3 \\
 & 4494.583 & 2.20 & $-$1.14 &      &   & 4.3 \\
 & 4528.605 & 2.18 & $-$0.82 & 13.8 &  4.62 & 4.3 \\
 & 4871.293 & 2.86 & $-$0.36 & 15.1 &  4.79 & \\
 & 4871.309 & 2.86 & $-$0.36 & 14.4 &  4.76 & \\
 & 4872.086 & 2.88 & $-$0.57 & 11.8 &  4.89 & \\
 & 4891.486 & 2.85 & $-$0.11 & 15.6 &  4.55 & \\
 & 4918.997 & 2.86 & $-$0.34 & 10.9 &  4.60 & \\
 & 4920.490 & 2.83 & \hspace {1.85ex}0.07 & 20.1 &  4.49 & 4.3 \\
 & 4957.594 & 2.81 & \hspace {1.85ex}0.23 & 27.9 &  4.52 & \\
 & 5192.367 & 3.00 & $-$0.42 &  7.6 &  4.61 & \\
 & 5216.422 & 1.61 & $-$2.15 &  2.1 &  4.53 & \\
 & 5227.206 & 1.56 & $-$1.23 & 14.6 &  4.48 & \\
 & 5232.945 & 2.94 & $-$0.06 & 13.0 &  4.47 & \\
 & 5232.946 & 2.94 & $-$0.06 & 15.3 &  4.55 & \\
 & 5269.530 & 0.86 & $-$1.32 & 45.0 &  4.72 & \\
 & 5269.537 & 0.86 & $-$1.32 & 51.7 &  4.88 & \\
 & 5270.321 & 1.61 & $-$1.34 & 23.7 &  4.91 & \\
 \hline
Fe\,II & 4233.160 & 2.57 & $-$2.00 & 16.9 &  4.73 & 4.3 \\
 & 4416.783 & 2.78 & $-$2.61 &  6.8 &  5.06 & \\
 & 4583.826 & 2.81 & $-$2.02 &  9.9 &  4.66 & \\
 & 4923.906 & 2.89 & $-$1.32 & 25.5 &  4.55 & 4.3 \\
 & 5018.431 & 2.89 & $-$1.22 & 33.6 &  4.64 & 4.3 \\
 & 5169.035 & 2.89 & $-$0.87 & 41.7 &  4.47 & 4.3 \\
 & 5275.868 & 3.20 & $-$1.94 &  4.5 &  4.52 & \\
\hline\hline
                \end  {tabular}
        \end  {flushleft}
\end  {table}

Although one of the star is a double lined spectroscopic binary,
similar methods can be used to determine the stellar parameters
 {\Teff} and  {\logg} of both stars.

\subsection  {Parameter determination methods} 

The temperature of the stars were determined from the profile of
 the wings of the hydrogen lines H$\alpha$ and H$\beta$. With
 \'echelle spectra, an accurate determination of the shape of the
 continuum above the H$\alpha$ wings is rather difficult. We used
 Gehren's method \citep {TG90} which gives an accuracy better
 than 0.5\,\% on the continuum placement. 
This corresponds to a random error of about 100\,K on the
 temperature of the star. 

\begin  {figure}[htbp]    
        \centering      
        \resizebox  {\hsize} {!}
         {\includegraphics  {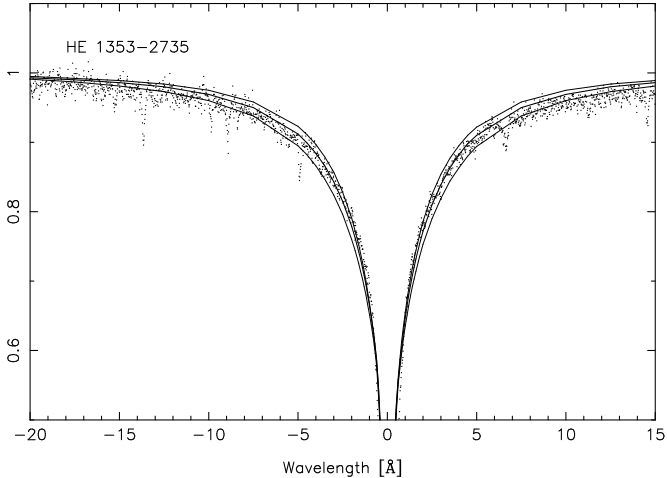} }  
        \caption  {The observed  {\Ha} wings of  {\dble} (dotted line) are
        used to constrain the  {\Teff} of the primary component.
        Overplotted are models computed for three  {\Teff}
        combinations for the binary system. From inner to outer curves:
        5700(A)+5100(B)\,K, 5900(A)+5200(B)\,K and 6100(A)+5500(B)\,K.}        
        \label  {F-Hadble}
\end  {figure}

Fig. \ref {F-Hadble} illustrates the sensitivity of temperature determination
using  {\Ha} in this case.

The models used in our analysis were interpolated in the grid of \cite {BE93},
computed with an updated version of the MARCS code \citep {BG75}, which
includes improved UV line blanketing \citep[see also][] {BE94}. These models,
and all calculations presented here, are fully LTE.

The gravity was determined by using the ionisation equilibrium balance of
Fe\,I and Fe\,II.  We caution that this method might yield gravities
systematically lower than the physical ones in very metal-poor turnoff stars
hotter than the sun; this phenomenon is being attributed to NLTE effects in
the atmosphere of metal-deficient stars \citep[see
e.g.][] {Fuhrmannetal:1997,thevenin99,allende99}.  However, no other method is
currently available for very metal-poor stars for which the distance is not
known. It should also be noted that NLTE effects act mainly on Fe\,I, whereas
Fe\,II shows a very small sensitivity to them. A good estimate of the true
metallicity of the stars can thus be achieved using Fe\,II lines.  In
addition, the gravities were checked on \cite {VDB97} isochrones.

We assumed a microturbulent velocity of 1.3\,\kms.  Note that the influence of
an error on the microturbulence parameter is negligible since all the metallic
lines are very weak ($<50$\,m {\AA}) and therefore hardly saturated.

\subsection {Specific methods for \dble}

Derivation of the individual components A and B and their corresponding
stellar parameters from the composite spectrum was performed iteratively. We
assume that the two components were born from the same gas cloud, and thus
have the same age and metallicity, and differ only by their mass.

To determine the temperature of both stars, we use the method
described in \cite {spite00}, which consists of the following steps. 
The temperature of the primary star is strongly constrained by the
profile of the wings of the hydrogen lines, while the temperature
difference between component A and B of the system is mainly 
constrained by the relative strength of the metallic lines of
both components.

Once the temperatures of the A and B components are chosen, the ratio of their
fluxes at any given wavelength are deduced from the bolometric magnitude of
the stars read on the theoretical HR diagram of \cite {VDB97}.  For this
purpose, the 12\,Gyr isochrone for $\mbox {[Fe/H]}=-2.3$\,dex was used, shifted
by 0.01 in $\log$\,\Teff, as needed to fit high parallax subdwarfs, following
\cite {CLPT97}. From the evolutionary tracks of \cite {VDB97} we
tentatively derive a mass ratio in the range $0.8<M_B/M_A<0.9$.

%
%
%

We note that it would be desirable to use isochrones with lower
metallicities than those provided by \cite {VDB97} in our analysis. However,
using different sets of isochrones has very little effect on the choice of
parameters, since only the  {\em relative flux\/} between the two components is
used, and this does not vary significantly from one set of isochrones to the
other. The resulting abundances are thus not sensitive to the choice of
isochrones, since the contribution of component B is a veil of the light
received from component A. 
The adopted atmospheric parameters are given in Table \ref {T-param}.
For comparison, parameters derived from the moderate-resolution
follow-up are also displayed, showing the very good agreement achieved.

\begin {table} 
  \caption {Adopted stellar parameters for the program
    stars. In the second row for each star we repeat the
    parameters derived from NTT/EMMI spectra for comparison.}\label {T-param}
  \begin {flushleft}
    \begin {tabular} {l@ {\hspace {2ex}}c@ {\hspace {1.5ex}}c@ {\hspace {1.5ex}}c@ {\hspace {1.5ex}}
        c@ {\hspace {1.5ex}}c@ {\hspace {1.5ex}}c}\hline\hline
      Star & \multicolumn {1} {l} {\Teff} & \logg &  [Fe/H] & \vt & Rel. flux & $M_ {bol}$
      \\[0.2ex]\hline\\[-2.2ex]
      \sgle    &  6500\,K & 4.2 &  $-$2.9  & 1.3 &      &     \\
               &  6500\,K & 4.2 &  $-$3.3  &     &      &     \\[0.5ex]
       {\dble}A &  5900\,K & 4.5 &  $-$3.2  & 1.3 & 1.00 & 5.52\\
               &  6000\,K & 4.8 &  $-$3.4  &     &      &     \\
       {\dble}B &  5200\,K & 4.5 &  $-$3.2  & 1.3 & 0.12 & 6.52\\\hline\hline
    \end {tabular}
  \end {flushleft}
\end {table}

\subsection  {Abundance determination}

\begin {figure}[htbp]    
  \centering      
  \resizebox  {\hsize} {!}
   {\includegraphics  {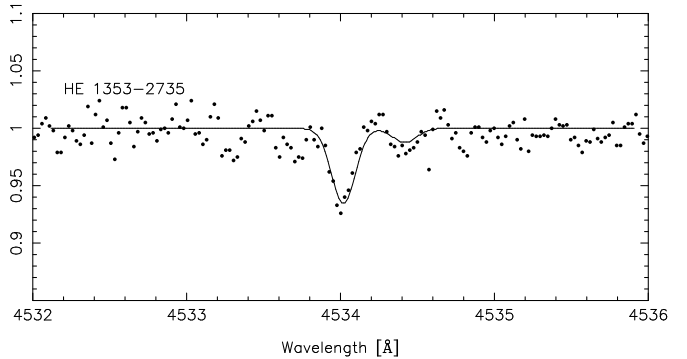} }     
  \caption  {Observed (dots) and synthetic (solid line) Ti\,II
  4534\, {\AA} line in \dble.}      
  \label {F-Tidble}
\end {figure}

For the single star (\sgle), abundances were determined by measuring the
equivalent widths by means of Gaussian fits of all the lines listed in Tables
\ref {T-lines} and \ref {T-linesFe}, from which individual abundances were then
computed using the model atmospheres of \cite {BE93}. For the binary (\dble),
each line was synthesised taking into account the contributions from the
components A and B. Fig. \ref {F-Tidble} is an example of such a synthetic
spectrum, overimposed on the observed spectrum.

For iron, the $\log gf$ values were taken from the compilation by \cite  {NJLTB94}.
The $\log gf$ values for Ti\,II are from \cite  {WF75} and
those for Mg\,I,  Ca\,I, Sc\,II and Sr\,II from \cite  {RNB91}.
The results from the computations are shown in Tables \ref {T-lines} and
\ref {T-linesFe}.

\section {Discussion}

\subsection {Stellar parameters}

\begin {table*}
\caption {Summary of the measured abundances.}\label {T-abundances}
\begin {center}
  \begin {tabular} {lrrrrrrrrrrr}\hline\hline
          & A(Li) & [Mg/H]  & [Ca/H]  & [Sc/H]  & [Ti/H]  & [Cr/H]  & [Fe/H]  & [Co/H]  & [Sr/H]  & [Y/H]   & [Ba/H]\\\hline
\sgle     & 2.15 & $-2.61$ & $-2.61$ & $-2.59$ & $-2.34$ & $-2.98$ & $-2.85$ & $-2.20$ & $-2.93$ & $-2.03$ & $-2.86$ \\
r.m.s     &      &   0.15  &   0.03  &   0.12  &   0.07  &   0.21  &   0.14  &         &         &   0.04  &         \\
$N$ lines &    2 &      4  &      5  &      3  &     12  &      5  &     52  &      1  &      1  &      1  &      2  \\[0.8ex]
\dble     & 2.06 &  $-2.8$ &  $-3.1$ &  $-3.4$ & $-2.90$  & $-3.47$ &  $-3.20$ &         & $<-4.4$ &         & $<-3.6$ \\  
r.m.s     &      &         &         &         &   0.15  &         &   0.10  &         &         &         &         \\
$N$ lines &    2 &      3  &      2  &      1  &      9  &       4 &     34  &         &      1  &         &       1 \\
\hline\hline
  \end {tabular}
\end {center}
\end {table*}

The iron content of the stars is $\sim 700\times$ and
$\sim 1600\times$ lower than solar ([Fe/H]=$-2.85$ and $-3.20$\,dex
for  {\sgle} and \dble, respectively). This is very close to what was
predicted from the HES and medium-resolution follow-up spectra.
However, \emph {both} stars show a slightly higher metallicity in the
high-resolution analysis. This might be due to the fact that 
the HES abundances were scaled to the HK survey scale by applying an
\emph {average} offset. Due to lack of data, it is currently not possible
to detect any \emph {trends} with [Fe/H] in the scaling, so that it is not
possible to exclude that the metallicities of the most metal-poor stars were
over-corrected. The effective temperatures and and surface gravities for both
stars are within the $1\,\sigma$ uncertainty of the parameters derived from
the moderate-resolution spectra.

\subsection {$\alpha$-elements}

As most stars in this metallicity range, both stars show
$\alpha$-elements enhancements, of the order of $+0.2$ to $+0.4$ with
respect to iron (Mg, Ca, Ti, see Table \ref {T-abundances}), which is
expected if the gas which gave birth to the stars has been enriched
preferentially by massive type\,II supernovae (SN\,II).

\subsection  {$s$-process elements} 

\begin {figure}[htbp]
  \begin {flushleft}
    \leavevmode
    \epsfig {file=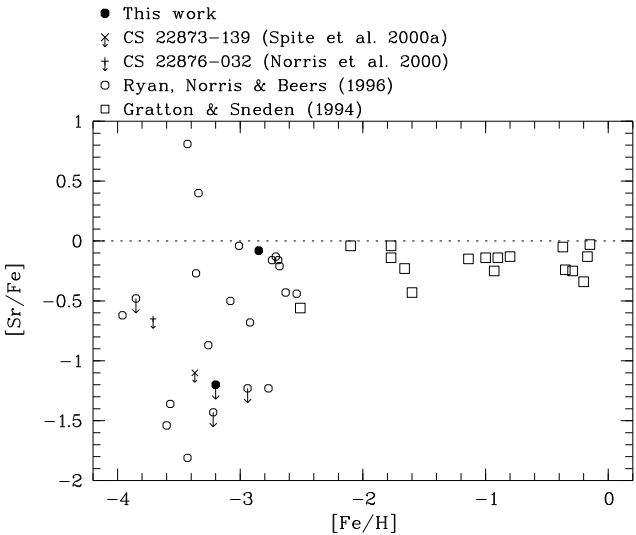, clip=, width=8.5cm,
      bbllx=68, bblly=425, bburx=370, bbury=681}
  \end {flushleft}
  \caption {\label {Fig:SrBaAbund} [Sr/Fe] as a function of [Fe/H]. Downward
    arrows refer to upper limits.}
\end {figure}

While Barium can also be synthesised via the $r$-process, Strontium and
Yttrium are thought to be synthesised solely via the ``slow neutron-capture''
process ($s$-process).
This process is thought to take place inside AGB stars, and given the
secondary nature of the process (iron-group elements need to be pre-existing
in the AGB), it starts enriching the Galaxy at later times than the products
of massive SN\,II. Indeed, at very low metallicities, all $s$-process elements
decline rapidly.  However, at metallicities lower than $-$2.5, the [Sr/Fe]
ratio becomes extremely scattered (see Fig. \ref {Fig:SrBaAbund}), by a factor
almost 200.  This large scatter is not found in other elements that can be
produced by the $s$-process: for example, no Ba-rich dwarf star ([Ba/Fe]
significantly larger than the average trend) has ever been detected in
low-metallicity stars, whereas Sr-rich objects are detected \citep[see e.g.
the review by][] {ryan00fs}.  The process at work is not yet elucidated: The
weak $s$-process in $\sim 15\,M_ {\sun}$ stars favors low-atomic number
neutron-capture species \citep {prantzos90} and would thus produce Sr without much of Ba, but
neutron sources at low metallicities would vanish and make the process
inefficient; a low neutron-exposure $r$-process \citep {ishimaru00} could be at work, but both
theoretical grounds and observational clues are lacking to conclude on this
topic.

Our program stars happen to sample this scatter in the [Sr/Fe] ratio,
 {\sgle} belonging to the Sr-normal group, and  {\dble} to the Sr-poor. 
It is interesting to note that  {\dble} is not the only Sr-poor
star which is in a binary system: CS\,22873-139 is a 19\,days period
extremely metal-poor binary with [Sr/Fe]$<-$1.1 \citep {spite00}, and
CS\,22876-032 (period 427\,days) is the most metal-poor dwarf known
to date, with  [Sr/Fe]$<-$0.65 \citep {norris00}.

\begin  {figure}[htbp]    
        \centering      
        \resizebox  {\hsize} {!}
         {\includegraphics  {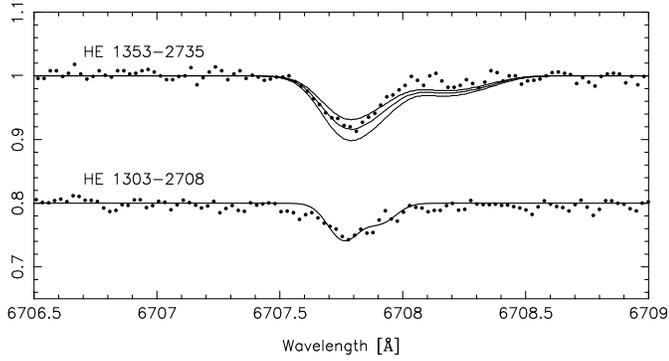}} 
        \caption  {Lithium 6708\, {\AA} line in the spectra of the two stars
        (dots). The straight lines are models with A(Li)=1.96, 2.06 and
        2.16 for  {\dble}, and A(Li)=2.15 for  {\sgle}. The spectrum of  {\sgle}
        was shifted vertically by $-$0.2 for the sake of clarity.}  
        \label {F-Liline}
\end  {figure}

\subsection {Lithium}

\begin {figure}[htbp]
  \begin {flushleft}      
    \resizebox {\hsize} {!} {\includegraphics {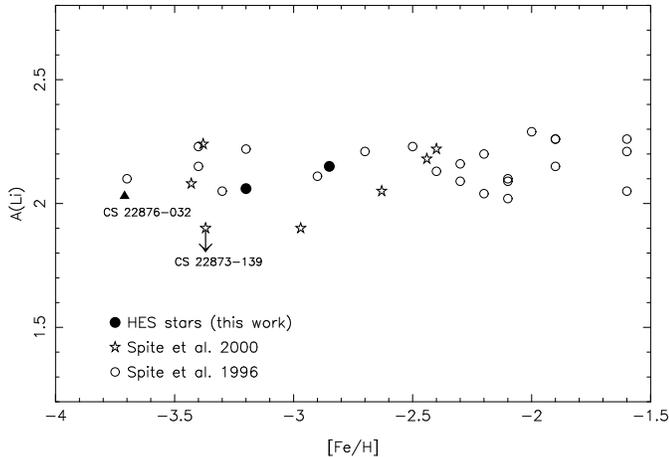}}       
    \caption {The Lithium-plateau for metal poor  
	dwarfs hotter than 5900\,K. The compilation is based on
	\cite {spite96}, complemented by \cite {spite00nat} for the
	most metal-poor stars, using the {\em same temperature scale}
      (based on {\Ha} wings). Also plotted are the two extremely
      metal-poor binaries CS\,22876-032 \citep {norris00} and
      CS\,22873-139 \citep {spite00}.}
    \label {F-LivFe}
  \end {flushleft}      
\end {figure}

In connection to the much debated primordial lithium abundance, and its
relation to the observed lithium abundance in metal-poor dwarfs (the so-called
Spite-plateau), we checked the status of this element in our two program
stars. We determined the abundance from the Li\,I\,6708\,{\AA} doublet by
comparing it to synthetic spectra.  Using the stellar parameters listed in
Table \ref {T-param}, we obtain A(Li)=2.15$\pm$0.05 and 2.06$\pm$0.1\,dex for
{\sgle} and {\dble}, respectively (see Fig. \ref {F-Liline}). The uncertainties
refer to the determination of the abundance itself, but do not include the
uncertainty arising from the choice of stellar parameters. In particular, the
Lithium abundance is very sensitive to effective temperature. We recall that
we used Balmer lines to determine \Teff, so that the present results should be
compared only with data making use of the {\em same {\Teff} scale}. For this
purpose, we plotted in Fig. \ref {F-LivFe} only data which used effective
temperatures determined from Balmer lines (see caption for details of
sources). As one can judge from the plot, both of our two stars are very close
to the average value of A(Li) for their metallicity.

\section {Conclusions}

We presented the first abundance analysis of metal-poor stars from the
Hamburg/ESO survey, based on high resolution spectra obtained with UVES at
VLT-UT2. With HE\,1353--2735, another Sr-rich, extremely metal-poor binary was
discovered.

The very good agreement of the stellar parameters deduced from
moderate-resolution, but good quality ($S/N> 50$) spectra with the parameters
derived from UVES spectra shows that the investment of considerable amounts of
observing time at 4\,m class telescopes pays off: reliably determined stellar
parameters ensure that no time is wasted at 8\,m class telescopes by observing
uninteresting objects.

From our investigation it is evident that much larger samples of extremely
metal-poor stars have to be analysed at high spectral resolution in order to
decipher the chemical history of our galaxy. With UVES at VLT-UT2 now being in
operation, this aim has become reachable. Fortunately there will be no lack of
metal-poor targets to be observed, since in the HES, hundreds of stars
at $\mbox {[Fe/H]}<-3.0$ are expected to be discovered, provided
moderate-resolution follow-up spectroscopy can be obtained for all
candidates. 

\begin {acknowledgements}
  We thank the Operations on Paranal and the UVES Science Verification team
  for the conduction of the observations and the timely public release of the
  data to the ESO community.  N.C. acknowledges financial support from
  Deutsche Forschungsgemeinschaft under grant Re~353/40, and acknowledges
  the hospitality shown to him at ESO headquarters.
\end {acknowledgements}

\bibliography {emps,mphs,ncpublications,HES}
\bibliographystyle {aabib99}

\end {document}